\documentstyle[11pt,newpasp,twoside,epsf]{article}
\def\edcomment#1{\iffalse\marginpar{\raggedright\sl#1\/}\else\relax\fi}
\reversemarginpar

\begin{document}
\title{On The Origin of {\sc Hi} in Galaxies: Photodissociation \\
and the ``Schmidt Law'' for Global Star Formation}
\author{Ronald J. Allen}
\affil{Space Telescope Science Institute,
3700 San Martin Drive, Baltimore, MD 21218, USA}

\begin{abstract} Young stars in the disks of galaxies produce {\sc Hi} from
their parent H$_2$ clouds by photodissociation.  This process is widespread
in late-type galaxies, and follows the distribution of Far-UV photons
produced primarily by B-type stars.  An estimate of the amount of
dissociated gas can be made using observed Far-UV fluxes and simple
approximations for the physics of photodissociation.  This leads to the
startling conclusion that much, and perhaps even all, of the HI in galaxy
disks can be produced in this way.  This result offers a simple, but
inverse, cause-effect explanation for the ``Schmidt Law'' of Global Star
Formation in galaxies.  \end{abstract}

\section{Introduction} \label{sec:intro}

Discussions on the subject of global star formation in galaxies (e.g.\
Kennicutt 1989, 1997, 1998; Martin \& Kennicutt 2001) are usually based on
observed correlations between measures of the gas surface density in galaxy
disks and the star formation rate.  An example of such a correlation
diagram is shown in Figure 1a (left panel), which plots on the X-axis the
mean surface brightness in the 21-cm line of {\sc Hi} for a sample of
local-universe galaxies, and on the Y-axis the corresponding observed mean
surface brightness at $\lambda = 1500$\AA\ taken from one of the first
extensive surveys of nearby galaxies in the Far-UV (the FAUST survey,
Deharveng et al.\ 1994).  Similar correlations appear if the data on the
Y-axis is the mean H$\alpha$ surface brightness of the galaxy, or if a
measure of the contribution from molecular gas (from CO(1-0) millimeter
line observations) is included on the X-axis (although this latter
``correction'' generally does not improve the degree of correlation, as was
first noted by Deharveng et al.).

\begin{figure}[t!]
\plottwo{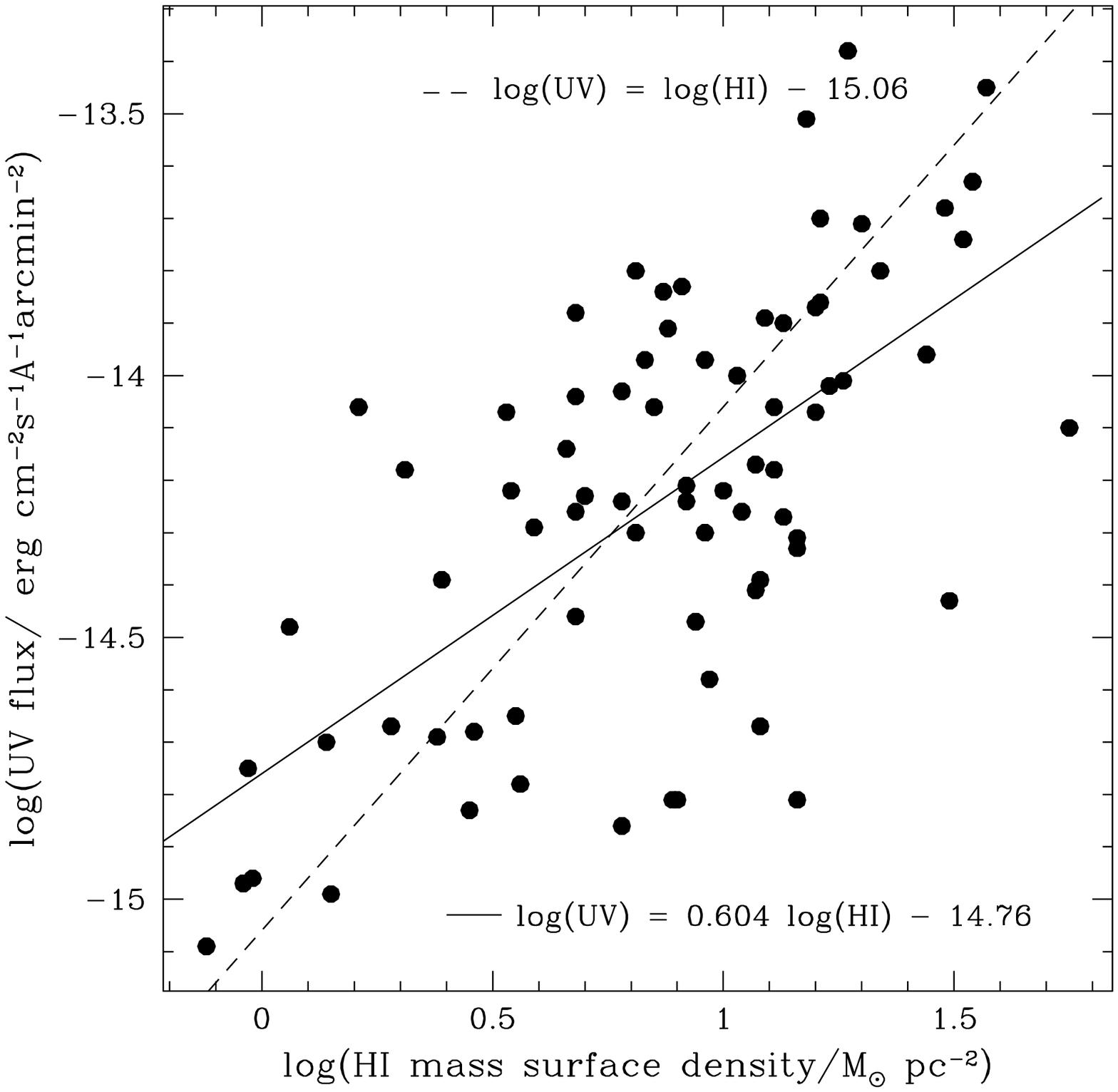}{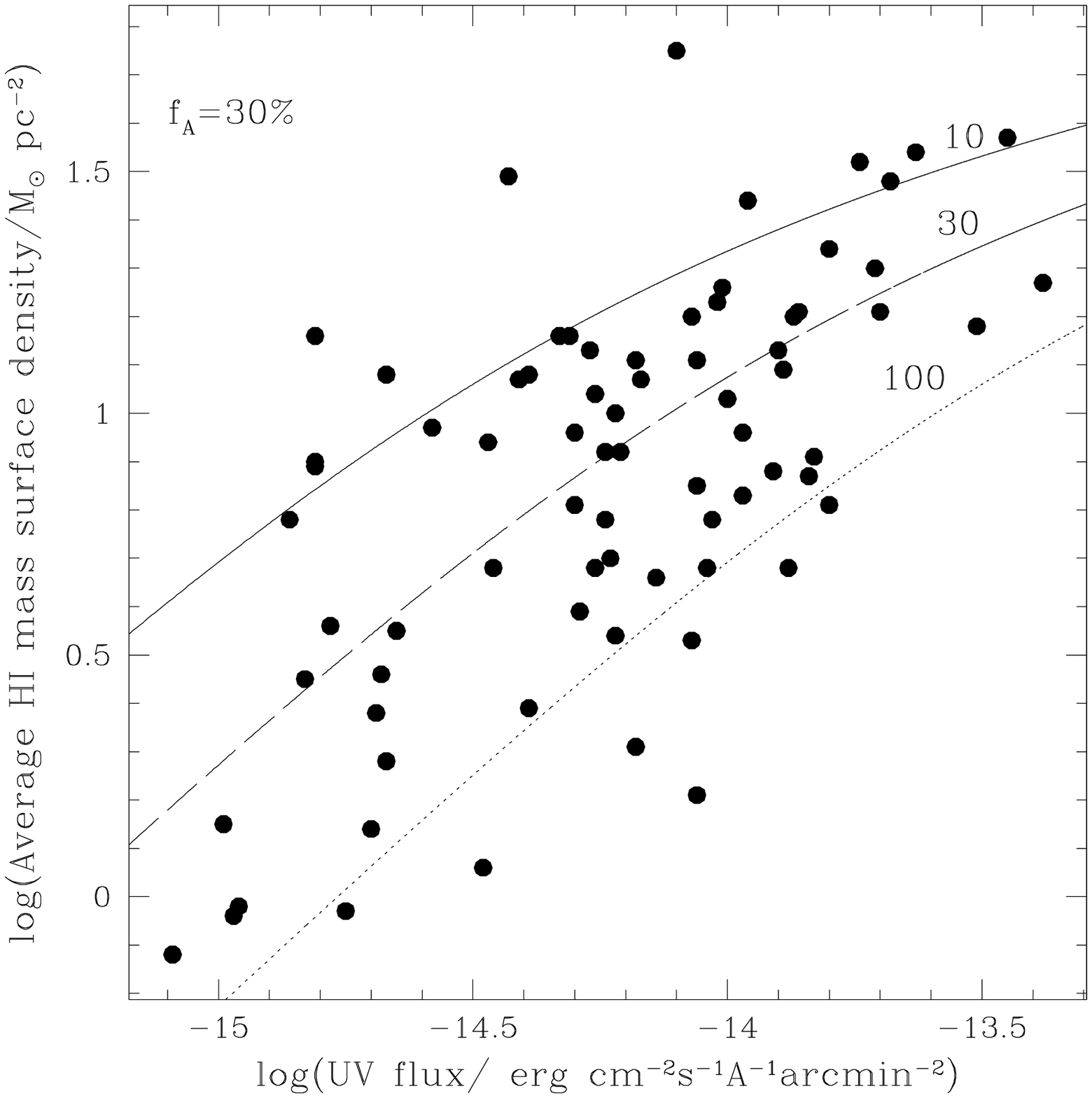}
\caption{ \textbf{a) Left Panel:}  Data showing a correlation between
average observed 21-cm line surface brightness (converted to ``{\sc Hi}
mass surface density'') on the X-axis and the Far-UV surface brightness
(taken as a measure of the formation rate of massive stars) on the Y-axis,
from Deharveng et al.\ (1994).  ``Schmidt Law'' fits are shown, with
indexes of 1 (dashed line) and 0.6 (solid line).  \textbf{b) Right Panel:}
The data plotted with axes inverted so as to emphasize the explanation in
terms of photodissociation.  The solid curves are the models of {\sc Hi}
production in PDRs described briefly in the text, and are labelled with the
proton volume densities of the parent GMCs.  The {\sc Hi} area filling
factor is assumed to be 0.30 over the disk of the galaxy.}
\end{figure}

\section{Cause, Effect, and the ``Schmidt Law''} \label{sec:schmidt}

The conventional approach is to parametrize plots such as that shown in
Figure 1a as a power law, following a suggestion by Schmidt (1959) who
assumed that the volume density of young stars formed per unit time in the
Galactic disk is proportional to some power of the local gas volume
density.  The same parametrization is usually assumed to apply to the
observed \textit{\textbf{surface}} densities in nearby galaxies (Kennicutt
1989).  Many authors have taken up the challenge to determine the power law
index for a variety of data sets and, some 38 years after Schmidt's seminal
paper, Kennicutt (1997) reported an accumulation of $\sim 50$ papers
dealing with the specific index appropriate for the ``Schmidt Law'' in
nearby galaxies.  The existence of such correlations is widely assumed to
be strong evidence that the original assumption by Schmidt is valid.

Inherent in nearly all discussions of this topic to date is another
assumption, namely, that the quantity plotted on X-axis of Figure 1a is the
\textit{cause}, and the quantity plotted on the Y-axis is the
\textit{effect}.  However, the mere existence of a correlation does not, of
course, tell us which quantity is the cause and which is the effect; it
only tells us that the two quantities are related.  Some other assumptions
are required to guide us in this matter; Schmidt's assumption provides this
guide.  It is against this backdrop that many papers have been published
discussing the global star formation rate in galaxies as a function of
various other parameters (galaxy type, galactocentric distance, etc.), and
the current picture of global instability in galaxy disks has been proposed
as the root cause of the star formation activity (Kennicutt 1989).

My purpose here is to point out that Figure 1 admits another explanation,
one that is simple, and for which the physics is well understood.  It is
well known that Far-UV photons are capable of dissociating H$_2$ molecules
into {\sc Hi} atoms in the ISM (Stecher \& Williams 1967), and that this
process creates {\sc Hi} layers on the surfaces of Giant Molecular Clouds
(GMCs) (e.g.\ Andersson \& Wannier 1993).  The dissociating photons are in
the energy range of a few eV below 13.6 eV and, since these photons are not
energetic enough to ionize the {\sc Hi}, they can travel over long
distances in the ISM, eventually being absorbed by an H$_2$ molecule or a
dust grain, or escaping the galaxy.  What has perhaps not been adequately
appreciated up to now is that the morphological signatures of
photodissociated {\sc Hi} can be found on a wide range of scales from 1 pc
to 1 kpc in the Galaxy and in nearby galaxies (Allen 2002).

Does this ``PDR'' explanation for {\sc Hi} in the ISM extend to whole
galaxies, and is the available flux of Far-UV photons from the typical
spiral galaxy in the local universe quantitatively sufficient to produce
the entire {\sc Hi} content of the galaxy for a reasonable range of
physical parameters?

\section{An Alternative Explanation for the ``Schmidt Law''}
\label{sec:explanation}

In Figure 1b, I have plotted the data from Deharveng et al.\ with the axes
inverted.  The solid curves are models of {\sc Hi} production by
photodissociation using the analytic solution developed by Sternberg (1988)
(see also Madden et al.\ 1993, Smith et al.\ 2000) for an {\sc Hi} area
filling factor of 0.3 and for a typical range of GMC volume densities $n =
n_1 + 2n_2$.  It appears from these results that photodissociation
is indeed capable of providing a simple explanation for the ``Schmidt Law''
in Figure 1a when we ``invert'' cause and effect, as in Figure 1b.  It is
also clear that the photodissociation picture can explain such correlations
when the H$\alpha$ flux replaces the Far-UV.

A very rough, but direct, order-of-magnitude estimate can be done to
confirm that the observed Far-UV flux provides enough photons to maintain
the entire {\sc Hi} content of the galaxy.  Consider the case of NGC 4152,
an Sc galaxy at D $\approx 19.5$ Mpc in Virgo.  The FAUST Far-UV flux is
$\approx 6.3 \times 10^{-14}$ ergs/cm$^2$/s/\AA\ integrated over the galaxy
image (this galaxy is just over $2'$ in diameter, so the corresponding data
point is roughly in the middle of Figure 1).  Summing over wavelength from
900 - 1100 \AA\ (and presuming the spectrum is flat from here to 1500 \AA)
the observed flux corresponds to about 1 photon/cm$^2$/s at earth, or
$\approx 4 \times 10^{52}$ Far-UV photons/s at NGC 4152.  What is the
appropriate time scale?  This will roughly be the reformation time scale on
dust grains for the process 2{\sc Hi} $\rightarrow$ H$_2$ on grains,
$t_{form} = (2nR_{form})^{-1} \approx 5 \times 10^8/n$ yr for the standard
PDR model parameters of Kaufman et al.\ (1999).  If we take a typical GMC
volume density to be $n_2 \approx 5 - 50$ H$_2$ molecules cm$^{-3}$, the
reformation time is typically $5 - 50 \times 10^6$ yr.  The {\sc Hi}
production rate by photodissociation in NGC 4152 then accounts for:

\[ 2 \times 0.15 \times f_t \times 4 \times 10^{52} \times 5 (50)
\times 10^6 \times 3.15 \times 10^7 \approx f_t \times 2 (20)
\times 10^{66}\:  {\rm
{\sc Hi}~atoms;}\]

\noindent where the factor 2 comes in because there are 2 {\sc Hi} atoms
produced by each Far-UV photon, the factor 0.15 is the dissociation
probability (calculated from atomic physics and averaged over all UV
absorptions, e.g.\ Draine \& Bertoldi 1996), and $f_t$ is the fraction of
the observed Far-UV flux which has been trapped in the galaxy and is
effective in causing photodissociation.

NGC 4152 is observed to contain $\approx 2 \times 10^9$ M$_{\odot}$ of {\sc
Hi} (Huchtmeier \& Richter 1989), or $2.5 \times 10^{66}$ {\sc Hi} atoms.
If a typical value for $f_t \approx 0.5$, then at least 40\%, and perhaps
\textit{\textbf{all}}, of the HI present could be accounted for by
photodissociation!  This result is surprising; a more complete discussion
will be presented elsewhere.

\section{Conclusions}

Correlations of the kind shown in Figure 1 can arise naturally from
the dissociating action of Far-UV photons reacting back on the GMCs from
which the young stars were recently formed.  Schmidt's hypothesis therefore
remains a reasonable, but unsubstantiated, point of departure for
discussions of the formation of stars out of the interstellar gas.

\acknowledgments
I am grateful to Ben Waghorn for his help with the figures, and to Nino
Panagia for discussions on photodissociation of H$_2$ and comments on an
earlier draft of this paper.

\end{document}